\documentclass[a4paper,12pt]{article}

\usepackage{amsmath}
\usepackage{hyperref}
\usepackage{cite}
\usepackage{amssymb}
\usepackage{braket}

\title{Off-shell Formalism for  Ali-Ilahi's ADHM Instanton Sigma Model}
\author{Abbas Ali\footnote{Email:aali.ph@amu.ac.in}, Mohsin Ilahi, P.P. Abdul Salih \\and Shafeeq Rahman Thottoli\footnote{Present Address : Department of Physical Sciences, Physics Division, College of Science,
Jazan University, Jazan 45142, Kingdom of Saudi Arabia}\\
		Physics Department, Aligarh Muslim University,\\ Aligarh, India}
\date{}
\begin{document}

\maketitle

\begin{abstract}
In this brief note we present the off-shell superspace formalism for Ali-Ilahi's $(0, 4)$ supersymmetric ADHM instanton linear sigma model in harmonic superspace. Ali-Ilahi's model is dual to the $(0, 4)$ supersymmetric ADHM instanton linear sigma model constructed by Witten in 1995.
\end{abstract}

Instantons, like solitons, are extended field theoretical solutions. A restricted class of Yang-Mills instantons was discovered by 't Hooft. Then there are general solutions constructed by Atiyah, Drinfeld, Hitchin and Manin (ADHM) \cite{Atiyah:1978ri}. We know a lot about 't Hooft's self-dual solutions like corresponding quantum corrections to the classical solutions \cite{tHooft:1976snw}. On the other hand the ADHM construction has been explored less, owing perhaps to its very brief nature of construction of a few pages. In 1995 constructed a sigma model incorporating the ADHM instanton \cite{Witten:1994tz}. This sigma model has $(0, 4)$ supersymmetry. This supersymmetry is in component and hence on-shell. We shall refer to it as the Original Model or Witten's original ADHM instanton linear sigma model. Galperin and Sokatchev constructed an off-shell version of Witten's ADHM instanton sigma model in harmonic superspace in Ref.\cite{Galperin:1994qn}. In Ref.\cite{Ali:2023csc} Ali and Ilahi constructed an ADHM instanton sigma model that is dual to the model constructed by Witten. We shall call it as the Complementary Model. The Complementary Model solved the quarter of a century old problem of the moduli space of the Original Model. Harmonic space off-shell formalism for the Complementary  Model was done in Ref.\cite{Ali:2025ntc}. In the present note we shall summarize the results of the last mentioned reference.

The duality between Witten's Original Model and Ali-Ilahi's Complementary Model is very simple in principle but cumbersome to implement. The duality involves the transformations : $(X, \psi)\leftrightarrow (\phi, \chi)$, $F=SU(2)\leftrightarrow F'=S(2)'$, $A\leftrightarrow A'$, $Y\leftrightarrow Y'$, $k\leftrightarrow k'$, $Sp(k)\leftrightarrow Sp(k')$ and $a\leftrightarrow a'$. Witten's original model has $F'=SU(2)'$ invariance while Ali-Ilahi's Complementary Model has $F=SU(2)$ invariance.

We begin by a brief summary of the harmonic superspace that is dual to the one used by Galperin and Sokatchev in above mentioned paper. This superspace is for $N=2$ supersymmetry. Next we collect the information for its generalization to the case where it can deal with the $(0, 4)$ supersymmetry of Ali-Ilahi's ADHM instanton sigma model. After that we summarize the free supermultiplets that are required for the off-shell superspace formalism of the Complementary Model. Then we describe the interactions of the same, that is, Ali-Ilahi's complementary linear ADHM instanton sigma model. After that the construction for the instanton gauge field is discussed. 
Finally in conclusions we shall discuss some results related to the subject matter matter of this note, that is, Ali-Ilahi's complementary ADHM instanton linear sigma model.

We begin by summarizing the  the harmonic superspace that is dual to the one used by Galperin and Sokatchev to formulate Witten's Original Model.

The harmonic variables for the harmonic superspace are denoted by $u^\pm_A$ and those of dual harmonic superspace by $\hat{u}^\pm_{A'}$ and the latter have the properties :
\begin{equation}\label{zweib0}
 \hat{u}^{+A'}\hat{u}^{-}_{A'}=1,~~~~~~   \begin{pmatrix}
\hat{u}^{-}_{1'} & \hat{u}^{+}_{1'} \\
\hat{u}^{-}_{2'} & \hat{u}^{+}_{2'} 
\end{pmatrix} \in SU(2)'.
\end{equation}
The dual harmonic variables are combined with dual Grassmann co-ordinates as
\begin{equation}\label{uone0}
\hat\theta^{+}_{\alpha}=\hat\theta^{A'}_{\alpha}\hat{u}^{+}_{A'}, ~~~~\hat\theta^{-}_{\alpha}=\hat\theta^{A'}_{\alpha}\hat{u}^{-}_{A'}
\end{equation}

The dual analytic subspace of the dual $N=2$ superspace looks like following:
\begin{equation}\label{subset0}
     \hat{x}^{\mu}_S = x^\mu-2i\hat\theta^{(A'}\sigma^\mu\bar{\hat\theta}^{B')}\hat{u}^{+}_{A'}u^{-}_{B'},~~~\hat\theta^{+}_{\alpha},~~~\bar{\hat\theta}^{+}_{\dot{\alpha}},~~~\hat{u}^{\pm}_{A'}.
\end{equation}
The dual $N=2$ supersymmetry transformations are 
\begin{eqnarray}\label{neq2susytr0}
 \delta\hat{x}^{\mu}_S &=&-2i(\hat\epsilon^{A'}\sigma^\mu\bar{\hat\theta}^{+}+\hat\theta^{+}\sigma^\mu\bar{\hat\epsilon}^{A'})\hat{u}^{-}_{A'},\nonumber\\
\delta\hat\theta^{+}_{\alpha}&=&\hat\epsilon^{A'}_{\alpha}\hat{u}^{+}_{A'},~~~\delta \bar{\hat\theta}^{+}_{\dot{\alpha}}=\bar{\hat\epsilon}^{A'}_{\dot{\alpha}}\hat{u}^{+}_{A'},~~~\delta \hat{u}^{\pm}_{A'}=0.    \end{eqnarray}

The superfield in the dual harmonic superspace corresponding to the superfield of original harmonic superspace  becomes:
\begin{eqnarray}\label{anasf0}
\hat\phi^{(q)}(\hat{x}_S,\hat\theta^+, \bar{\hat\theta}^+, \hat{u}^\pm)&=&\hat{F}^{(q)}(\hat{x}_S, \hat{u}^{\pm})+\hat\theta^{\alpha+}\hat\psi^{(q-1)}_{\alpha}(\hat{x}_S, \hat{u}^{\pm})\nonumber \\
&+&\bar{\hat\theta}^{+}_{\dot{\alpha}}\bar{\hat\varphi}^{\dot{\alpha}(q-1)}(\hat{x}_S, \hat{u}^{\pm})
+\bar\theta^{+}\bar\theta^{+}\hat{M}^{(q-2)}(\hat{x}_S, \hat{u}^{\pm})\nonumber \\
&+&\bar{\hat\theta}^{-}\bar{\hat\theta}^{+}\bar{N}^{(q-2)}(\hat{x}_S, \hat{u}^{\pm}) + \hat\theta^{+}\sigma^a\bar{\hat\theta}^{+}\hat{A}_{a}^{(q-2)}(\hat{x}_S, \hat{u}^{\pm}) \nonumber \\
&+& \bar{\hat\theta}^{+}\bar{\hat\theta}^{+}{\hat\theta}^{\alpha+}\hat{\xi}^{q-3}_{\alpha}(\hat{x}_S, u^{\pm})+\hat\theta^+\hat\theta^+\bar{\hat\theta}^+_{\dot{\alpha}}\bar{\hat\chi}^{\dot{\alpha}(q-3)}(\hat{x}_S, \hat{u}^{\pm}) \nonumber \\
&+& \hat\theta^+\hat\theta^+\bar{\hat\theta}^+\bar{\theta}^{+}\hat{D}^{(q-4)}          (\hat{x}_S, \hat{u}^{\pm}).
\end{eqnarray}
Here we are using the notation $q$ for $U(1)'$. The advantage of the analytic subspace is that it greatly reduces the number field components of the superfield.

The three derivative operators in dual harmonic superspace preserving the reality conditions on harmonic variables are given by 
\begin{equation}\label{threedefop0}
\hat{D}^{++} = \hat{u}^{+A'}\frac{\partial}{\partial \hat{u}^{-A'}},~~~    \hat{D}^{--} = \overline{\hat{D}^{++}},~~~ \hat{D}^{0} = \hat{u}^{+A'}\frac{\partial}{\partial \hat{u}^{+A'}} - \hat{u}^{-A'}\frac{\partial}{\partial\hat{u}^{-A'}}. 
\end{equation}
These obey the $SU(2)'$ algebra.

 In the dual harmonic superspace he  eigen-functions of the charge operator are defined by the following equation.

\begin{equation}\label{harmfn0}
\hat{D}^{0}\hat{f}^{q}(\hat{u}) = q\hat{f}^{q}(\hat{u}) ,~~~ q=0,\pm 1, \pm 2,\cdots.
\end{equation}
The harmonic expansion of the related field is given by:
\begin{equation}\label{harmexp0}
\hat{f}^{q}({u})=\sum_{n=0}^{\infty}\hat{f}^{A'_1\cdots A'_{n+q}B'_1\cdots B'_n}\hat{u}^{+}_{(A_1}\cdots \hat{u}^{+}_{A'_{n+q}}\hat{u}^{-}_{B'_1}\cdots\hat{u}^{-}_{B'_n)}.
\end{equation}

 For dual harmonic superspace the reality conditions on harmonic functions are
 \begin{equation}\label{partconj0}
\widetilde{\hat{f}^{A'B'\cdots}}=\overline{\hat{f}^{A'B'\cdots}},~~~\widetilde{\hat{u}^{\pm {A'}}}=\hat{u}^{\pm}_{A'},~~~\widetilde{\hat{u}^{\pm}_{A'}}=-\hat{u}^{\pm A'}.
 \end{equation}

 The rules for integration in the dual harmonic superspace are:
\begin{equation}\label{harmint1}
\int d\hat{u} 1=1,~~\int d\hat{u} \hat{u}^{+}_{(A'_1}\cdots\hat{u}^{+}_{A'_{p}}\hat{u}^{-}_{B_1}\cdots\hat{u}^{-}_{B_r)} ~~~\text{for} ~p+r>0.
\end{equation}

In the expansion of the integrand, the dual harmonic integral essentially projects out the singlet part,  just like the original harmonic superspace.   We can also prove the following identity:
\begin{equation}\label{identity0}
\hat{D}^{++}_1 \frac{1}{\hat{u}^{+}_1\hat{u}^{+}_2}=\delta^{+,-}(\hat{u}_1,\hat{u}_2),
\end{equation}
where $\hat{u}^{+}_1\hat{u}^{+}_2 \equiv \hat{u}^{+A'}_1\hat{u}^{+}_{2A'}$ and $\delta^{+,-}(\hat{u}_1, \hat{u}_2)$ is a harmonic delta function. This is   equivalent to
\begin{equation}\label{delta-rule0}
    \frac{\partial}{\partial \bar z}=\pi\delta(z).
\end{equation}

We next summarize a few aspects of Ali-Ilahi's Complementary Model in the spirit of Galperin and Sokatchev.

This dual or complementary ADHM construction uses the tensors $A^{a'}_{A'Y}$ with  $SO(n')$  index $a' = 1, \cdots, n'+4k$ and  $Y = 1, \cdots, 2k$ (where $n'$ and $k$ are positive numbers) and $A'$ being the $SU(2)'$ index.

There is the following reality condition on this tensor :
\begin{equation}\label{real3}
\overline{A^{a'}_{A'Y}}=\epsilon^{A'B'}\epsilon^{YZ}A^{a'}_{B'Z}.
\end{equation}
This tensor $A^{a'}_{A'Y}(\hat\phi)$ is linear in $\hat\phi$ 
\begin{equation}
 A^{a'}_{A'Y}(\hat\phi)=\hat N^{a'}_{A'Y}+\hat E^{a'}_{YY'}\hat\phi_{A'}^{Y'}\label{linear5}   
\end{equation}
where $\hat N^{a'}_{A'Y}$ and $\hat E^{a'}_{YY'}$ are constants. It must satisfy the algebraic constraint:
\begin{equation}\label{real5}
A^{a'}_{A'Y}(\hat\phi)A^{a'}_{B'Z}(\hat\phi)=\epsilon^{A'B'}\hat R_{YZ}(\hat\phi),
\end{equation}
here $\hat R_{YZ}$ is an invertible antisymmetric $2k\times 2k$ matrix.

For successful implementation of the ADHM construction the matrices $\hat N^{a'}_{A'Y}$ and $\hat  E^{a'}_{YY'}$ should have maximal rank. 

The harmonic superspace formulated by Galperin and the corresponding dual harmonic superspace constructed above are both for $N=2$ supersymmetry. The latter one has to be generalized so that it can deal with the $(0, 4)$ supersymmetry of Ali-Ilahi's ADHM instanton sigma model. This is what we take up now.

We begin with the analytic basis for the enlarged dual harmonic superspace:
\begin{equation}\label{bas0}
\hat{x}_{S++} = x_{++} + i\hat{\theta}^{AA'}_+\hat{\theta}^{B'}_{+A} \hat{u}^+_{(A'}\hat{u}^-_{B')},
\ \ x_{--}, \ \ \hat{\theta}^{\pm A}_{+}= \hat{u}^\pm_{A'} \hat{\theta}^{AA'}_+, \ \ \hat{u}^\pm.\end{equation}
The derivative $\hat{D}^+_{-A'}$ in the dual harmonic space is defined as :
\begin{equation}
    \hat{D}^+_{-A} =
\frac{\partial}{\partial\hat\theta^{-A}_+}.
\end{equation}
The Grassmann analyticity criterion becomes:
\begin{equation}\label{asf0}
\hat{D}^+_{-A}\hat\Phi(\hat{x},{\hat\theta},\hat{u}) = 0 \ \ \Rightarrow \ \ \hat\Phi = \hat\Phi(\hat{x}_{S++},
x_{--}, {\hat\theta}^+_+, \hat{u})\; .
\end{equation}
The dual harmonic derivative ${D}^{++}$ receives a vielbein term and becomes:
\begin{equation}\label{der0}
\hat{D}^{++} = \hat{u}^{+A}\frac{\partial}{\partial\hat{u}^{-A'}} + i{\hat\theta}^{+A'}_+{\hat\theta}^+_{+A'}
\frac{\partial}{\partial\hat{x}_{S++}}.
\end{equation}
Thus we have
\begin{equation}\label{hgr0}
[\hat{D}^{++}, \hat{D}^+_{-A}] = 0.
\end{equation}
 
The analytic superfields  have a non-vanishing $U(1)$ harmonic charge $q$ short Grassmann expansion,
\begin{equation}\label{expa0}
\hat\Phi^q(\hat{x},{\hat\theta}^+,\hat{u}) = \hat\phi^q(\hat{x}, \hat{u}) + {\hat\theta}^{+A}_+\hat\xi^{q-1}_{-A'}(\hat{x}, \hat{u})
+({\hat\theta}^+_+)^2 \hat{f}^{q-2}_{--}(\hat{x}, \hat{u})\; ,
\end{equation}
where $({\theta}^+_+)^2 \equiv {\theta}^{+A}_+{\theta}^+_{+A}\;$. The coefficients in Eqn.(\ref{expa0}) are harmonic-dependent fields (remember that all of the terms in Eqn.(\ref{expa0}) conserve the total $U(1)$ charge $q$).  

Witten's Original Model uses three supermultiplets : fundamental scalar field $(X, \psi)$, chiral supermultiplet $(\lambda^a, F^a)$  and the twisted scalar multiplet $(\phi, \chi)$. Since Ali-Ilahi's complementary ADHM instanton linear sigma model is dual to the Witten's Original Model this too needs three supermultiplets. These are : the fundamental scalar supermultiplet $(\hat\phi, \hat\psi)$, the chiral supermultiplet $(\hat\lambda^a, \hat{F}^{a'})$ and the twisted scalar supermultiplet $(\hat{X}, \hat\psi)$. For off-shell formulation these are needed in the dual harmonic superspace. We now collect information about these now. 

We begin with the construction of the dual scalar superfield based on scalar supermultiplet $(\hat\phi, \hat\chi)$. Here $\phi^{A'Y'}$ are the coordinates of the target space $R^{4k'}$ because $A'$ takes two values in $SU(2)'$ and $Y'$ takes $2k'$ values in $Sp(k')$. Corresponding supermultiplet in harmonic superspace is defined as :

 \begin{equation}\label{X0}
\hat\Phi^{+Y'}(x, {\hat\theta}^+, \hat{u}) = \hat\phi^{+Y'}(x, \hat{u}) + i{\hat\theta}^{+A}_+\hat\chi^{Y'}_{-A}(x, \hat{u})
+({\hat\theta}^+_+)^2 \hat{f}^{-Y'}_{--}(x, \hat{u}).
\end{equation}

These superfields are real in the following sense:
\begin{equation}\label{Xreal0}
\widetilde{\hat\Phi^{+Y'}} = \epsilon_{Y'Z'}\hat\Phi^{+Z'}.
\end{equation}

Corresponding superfield action is
\begin{equation}\label{acX0}
\hat{S}_{\hat\Phi} = i\int d^2x d\hat{u}d^2{\hat\theta}^+_+ \hat\Phi^{+Y'}\partial_{++} \hat\Phi^+_{Y'}.
\end{equation}
The component action is:
\begin{equation}\label{comX0}
\hat{S}_{\hat\phi} = \int d^2x\; \left( \hat\phi^{A'Y'}\partial_{++}\partial_{--}\hat\phi_{A'Y'} +
\frac{i}{2}\hat\chi^{AY'}_-\partial_{++}
\hat\chi_{-AY'}\right).
\end{equation}
We now come to the second supermultiplet that is a chiral fermion supermultiplet.
We take the following real superfield that are {\it anti-commuting}:
\begin{equation}\label{Lam0}
\hat{\Lambda}^{a'}_+(x,\hat{\theta}^+,\hat{u}) = \hat\lambda^{a'}_+(x,\hat{u}) + \hat{\theta}^{+A}_+\hat{g}^{-a'}_{A}(x,\hat{u})
+i(\hat{\theta}^+_+)^2  \hat\sigma^{--a'}_{-}(x,\hat{u}).
\end{equation}
Corresponding superfield action is
\begin{equation}\label{acL0}
S_{\hat\Lambda} = \frac{1}{2}\int d^2x d\hat{u} d^2\hat{\theta}^+_+ \;
\hat\Lambda^{a'}_+\hat{D}^{++} \hat\Lambda^{a'}_{+}.
\end{equation}
In component form it becomes:
\begin{equation}\label{chfe0}
S_{\hat\lambda} = \frac{i}{2}\int d^2x \;
\hat\lambda^{a'}_+(x)\partial_{--}\hat\lambda^{a'}_{+}(x).
\end{equation}

Finally we take up the third supermultiplet that is the twisted scalar supermultiplet based on the fields $\hat{X}^{Y}_{A}(x,\hat{u})$:
\begin{equation}\label{X1}
\hat X^{+Y}_+(x,\hat{\theta}^+,\hat{u}) = \hat\rho^{+Y}_+(x,\hat{u}) +
\theta^{+A}_+\hat X^{Y}_{A}(x,\hat{u}) +i(\hat{\theta}^+_+)^2 \hat\psi^{-Y}_{-}(x,\hat{u}).
\end{equation}
Corresponding superfield action is
\begin{equation}\label{acP0}
\hat{S}_{\hat X}=\int d^2x d^4\hat{\theta}_+ d\hat{u}_1d\hat{u}_2\; \frac{1}{\hat{u}^+_1\hat{u}^+_2}
\hat X^{+Y}_+(1)\partial_{++} \hat X^+_{+Y}(2).
\end{equation}

Corresponding component action is
\begin{equation}\label{comP0}
S_{\hat X} = \int d^2x\; \left( \hat{X}^{YA}\partial_{++}\partial_{--}
\hat{X}_{YA} + \frac{i}{2}\hat\psi^{YA'}_-
\partial_{++}\hat\psi_{-YA'} \right).
\end{equation}

We now take up the description of  the interactions of Ali-Ilahi's complementary ADHM instanton linear sigma model. 

The dual ADHM interaction  takes the following form:
\begin{equation}\label{int0}
\hat{S}_{int} = m \int d^2x d\hat{u} d^2\hat{\theta}^+_+ \;  \hat {X}^{+Y}_+ \hat{v}^{+a'}_{Y}(\hat\Phi^+,\hat{u})
\hat\Lambda_+^a,
\end{equation}
with
\begin{equation}\label{c10}
\hat{v}^{+a'}_{Y}\hat{v}^{+a'}_{Z} = 0,
\end{equation}
\begin{equation}\label{c20}
\hat{D}^{++}\hat{v}^{+a'}_{Y}(\hat\Phi^+,\hat{u}) = 0.
\end{equation}
\begin{equation}\label{lin0}
\hat{v}^{+a'}_{Y}(\hat\Phi^+,\hat{u}) = \hat{u}^{+A'} \hat{N}^{a'}_{A'Y} + \hat{E}^{a'}_{YY'}\hat\Phi^{+Y'},
\end{equation}
\begin{equation}\label{lin'0}
\hat{v}^{+a'}_{Y}(\hat\Phi^+,\hat{u})|_{\hat{\theta}=0} = \hat{u}^{+A'} (\hat{N}^{a'}_{A'Y}  +
\hat{E}^{a'}_{YY'} \hat\phi^{Y'}_{A'}) \equiv \hat{u}^{+A'}
A^{a'}_{A'Y}(\hat\phi).
\end{equation}
Where we have defined
\begin{equation}
 \hat{N}^{a'}_{A'Y}  +
\hat{E}^{a'}_{YY'} \hat\phi^{Y'}_{A'} = A^{a'}_{A'Y}(\hat\phi).  \label{linear6}
\end{equation}
Here in Eqn.(\ref{linear6}) we have recovered the Eqn.(\ref{linear5}).

The last element in the construction of Ali-Ilahi's complementary ADHM instanton linear sigma model is the construction of the instanton gauge field. This is what we take up now.

The superfield form of the Lagrangian for the dual ADHM instanton is given by the following expression
\begin{equation}\label{ADga0}
\hat{\cal L}^{++}_{++}(\hat\Lambda)|_{m\rightarrow\infty} = \hat\Lambda^{i'}_+[\delta^{i'j'}
\hat{D}^{++} +
(\hat{V}^{++})^{i'j'}] \hat\Lambda^{j'}_+,
\end{equation}
where
\begin{equation}\label{calV0}
(\hat{V}^{++})^{i'j'} = \hat{v}^{i'a'}(\Phi^+,\hat{u})\hat{D}^{++}\hat{v}^{j'a'}(X^+,\hat{u}).
\end{equation}
Here the   $2k\times (n'+4k)$ matrix
$\hat{v}^{+a'}_{Y}(\Phi^+,\hat{u})$ has been promoted to
full {\it orthogonal} matrix $\hat{v}^{{\tilde a}'a'}(\Phi^+,\hat{u})$, where the $n'+4k$
dimensional index ${\tilde a} = (+Y, -Y, i')$ and
$i'=1,\cdots, n'$ is an index of the group $SO(n')$. This time orthogonality
means the following
\begin{equation}\label{or0}
v^{\tilde a'a'} v^{\tilde b' a'} = \delta^{\tilde a'\tilde b'}.
\end{equation}
Which, in component form, gives
\begin{equation}\label{ferr0}
\hat{S} = \int d^2x \; \left( \frac{i}{2}\hat\lambda^{a'}_+\partial_{--}\hat\lambda^{a'}_{+}
- \frac{m}{2} \hat\psi^{A'Y}_- A^{a'}_{A'Y}(\hat\phi)\hat\lambda^{a'}_+\right).
\end{equation}

This leads to the following explicit form for the dual instanton:
\begin{equation}\label{fergau0}
\hat{S} =\frac{i}{2} \int d^2x \; \hat\lambda^{i'}_+(\delta^{i'j'}\partial_{--} +
\partial_{--}\hat\phi^{A'Y'} \hat{A}^{i'j'}_{A'Y'})
\hat\lambda^{j'}_+,
\end{equation}
where
\begin{equation}\label{ADHMgau0}
\hat{A}^{i'j'}_{A'Y'} = \hat{v}^{i'a'} \frac{\partial\hat{v}^{j'a'}}{\partial \hat\phi^{A'Y'}}.
\end{equation}

We shall now take the results related to the subject matter of this note.

As stated above Witten's Original Model ( as well as Ali-Ilahi's Complementary Model) have $(0, 4)$ supersymmetry. This has an $SO(4)$ automorphism. In both constructions, Original and Complementary, this automorphism breaks as $SO(4)\sim SU(2)\times SU(2)$. Witten names these $SU(2)$'s as $SU(2)=F$ and $SU(2)'=F'$. Witten's construction preserved the $F'$ symmetry and broke $F$. Ali-Ilahi's model preserved $F$ and broke $F'$. The suggestion for the complementary construction was made by Witten himself. He also suggested to keep both of the $F$ and $F'$ symmetries in a construction. This was done in Ref. \cite{Ali:2023icn} by Ali and Salih. This was called the Complete Construction, in line with Original and Complementary constructions.

Quantization of Witten's original model was done by Lambert in Ref. \cite{Lambert:1995dp}. On the basis of this the quantization of the Complementary Model was done by Ali, Ilahi and Thottoli in Ref.\cite{Ali:2023zxt}

In Ref.\cite{Giveon:1998ns} Giveon, Kutasov and Seiberg made very perceptive comments on $AdS_3$ superstrings. In their investigations they encountered a mysterious doubling of the Ramond superalgebra in case of $AdS_3$ superstrings. This problem, like the problem of the moduli space of Witten's Original Model, remained unsolved for quarter of a century. On the basis of insights from the Complementary Model this problem was solved in Ref. \cite{Ali:2023xov}.

In Ref.\cite{Gukov:2004ym} Gukov, Martinec, Moore and Strominger (GMMS) examined the proposed duals to superstrings moving on $AdS_3\times S^3\times S^3\times S^1$ background and rejected all of the then existent proposals. Since then there has been important advancements in this regard but the problem remains essentially unresolved. GMMS tried to extract insights about the problem by taking the tensor product of two Sevrin, Troost and van Proeyen's free field realization the large $N=4$ superconformal algebra \cite{Sevrin:1988ew,  Ali:2000zu, Ali:2003aa, Ali:2000we, Ali:1993sd, Ivanov:1987mz, Ivanov:1988rt}. They gave up their endeavour because of the non-associativity of the tensor product. In Ref. \cite{Ali:2023kkf} Ali and Ilahi solved this two decades old problem because of the $Z_2$ symmetry that exists between the Original Model and the Complementary Model and that is inherited by the GMMS problem in the individual, not general, case.

The symmetry structure of the Original Model, the Complementary Model and the Complete Model are reflected in the symmetry structure of superstrings on $AdS_3\times S^3\times K3$, $AdS_3\times S^3\times T^4$ and $AdS_3\times S^3\times S^3\times S^1$ backgrounds. This was demonstrated by Ali, Ilahi, Salih and Thottoli in Ref.\cite{Ali:2024amc}. 

In Ref.\cite{Banados:1992wn} the authors constructed a three dimensional rotating black hole that has become famous at the BTZ solution. Among other people Ali and Kumar constructed an stringy version in Ref.\cite{Ali:1992mj}. It has come to our notice that when people talk about BTZ solution in the context of string theory they still use the former solution and not the latter. Clearly this situation should change. We would like to point out that in view of the close connection between BTZ black hole and $AdS_3$ superstrings and the connection of the ADHM instanton linear sigma models further investigations should reveal interesting information.

{\it Acknowledgment :} AA and MI thank Professor DP Jatkar and Professor R. Gopakumar for hospitality. AA thanks Professor V. Ravindran for hospitality. MI, PPAS and SRT contributed to this work during their PhD work.


\begin{thebibliography}{99}

\bibitem{Atiyah:1978ri}
M.~F.~Atiyah, N.~J.~Hitchin, V.~G.~Drinfeld and Y.~I.~Manin,
``Construction of Instantons,''
Phys. Lett. A \textbf{65} (1978) 185-187

\bibitem{tHooft:1976snw}
G.~'t Hooft,
``Computation of the Quantum Effects Due to a Four-Dimensional Pseudoparticle,''
Phys. Rev. D \textbf{14}, 3432-3450 (1976)
[erratum: Phys. Rev. D \textbf{18}, 2199 (1978)]

\bibitem{Witten:1994tz}
E.~Witten,
``Sigma models and the ADHM construction of instantons,''
J. Geom. Phys. \textbf{15} (1995) 215-226
[arXiv:hep-th/9410052].

\bibitem{Galperin:1994qn}
A.~Galperin and E.~Sokatchev,
``Manifest supersymmetry and the ADHM construction of instantons,''
Nucl. Phys. B \textbf{452} (1995) 431-455
[arXiv:hep-th/9412032 [hep-th]].

\bibitem{Ali:2023csc}
A.~Ali and M.~Ilahi,
``Complementary ADHM Instanton Sigma Model,''
[arXiv:2305.05951 [hep-th]].

\bibitem{Ali:2025ntc}
A.~Ali, M.~Ilahi, P.~P.~A.~Salih and S.~R.~Thottoli,
``Harmonic Superspace for Ali-Ilahi's ADHM Instanton Sigma Model,''
[arXiv:2507.22948 [hep-th]].

\bibitem{Ali:2023icn}
A.~Ali and P.~P.~A.~Salih,
``Complete ADHM Sigma Model,''
[arXiv:2305.09516 [hep-th]].

\bibitem{Lambert:1995dp} 
N.~D.~Lambert,
``Quantizing the (0,4) Supersymmetric ADHM Sigma Model,''
Nucl. Phys. B \textbf{460} (1996), 221-232
[arXiv:hep-th/9508039].

\bibitem{Ali:2023zxt}
A.~Ali, M.~Ilahi and S.~R.~Thottoli,
``Quantization of Complementary ADHM Sigma Model,''
[arXiv:2306.10002 [hep-th]].

\bibitem{Giveon:1998ns}
A.~Giveon, D.~Kutasov and N.~Seiberg,
``Comments on string theory on $AdS_3$,''
Adv. Theor. Math. Phys. \textbf{2}, 733-782 (1998)
[arXiv:hep-th/9806194 [hep-th]].

\bibitem{Ali:2023xov}
A.~Ali,
``Superalgebra Doubling in $AdS_3$ Superstrings,''
[arXiv:2306.11047 [hep-th]].

\bibitem{Gukov:2004ym}
S.~Gukov, E.~Martinec, G.~W.~Moore and A.~Strominger,
``The Search for a holographic dual to $AdS(3) \times S^3 \times ^3 \times S^1$,''
Adv. Theor. Math. Phys. \textbf{9}, 435-525 (2005)
[arXiv:hep-th/0403090 [hep-th]].

\bibitem{Sevrin:1988ew} 
A. Sevrin, W. Troost and A. van Proeyen, ``Superconformal Algebras in Two-Dimensions with N=4'', Phys. Lett. B 208 (1988) 447.
 

\bibitem{Ali:2000zu}
A.~Ali,``Free Field Realizations of N=4 Superconformal Algebras,''
Indian J. Pure Appl. Phys. \textbf{38} (2000) 446-452
			
\bibitem{Ali:2003aa}
A.~Ali,``Types of Two-dimensional N = 4 Superconformal Field Theories,''
Pramana \textbf{61} (2003) 1065-1078
[arXiv:hep-th/9906096].

\bibitem{Ali:2000we}  
A.~Ali,``Conformal Symmetry of Superstrings on $AdS_3 \times S^3 \times T^4$ and D1 / D5 system,''
Mod.\ Phys.\ Lett.\ A {\bf 17} (2002) 2477
[hep-th/0007021].

\bibitem{Ali:1993sd}
A.~Ali and A.~Kumar,
``A New N=4 superconformal algebra,''
Mod. Phys. Lett. A \textbf{8}, 1527-1532 (1993)
[arXiv:hep-th/9301010 [hep-th]].

\bibitem{Ivanov:1987mz}
E.~A.~Ivanov, S.~O.~Krivonos and V.~M.~Leviant,
``A New Class of Superconformal $\sigma$ Models With the {Wess-Zumino} Action,''
Nucl. Phys. B \textbf{304} (1988), 601-627

\bibitem{Ivanov:1988rt} 
E.~A.~Ivanov, S.~O.~Krivonos and V.~M.~Leviant,
``Quantum N=3, N=4 Superconformal WZW Sigma Models,''
Phys. Lett. B \textbf{215} (1988), 689
[erratum: Phys. Lett. B \textbf{221} (1989), 432]


\bibitem{Ali:2023kkf}
A.~Ali and M.~Ilahi,
``$Z_2$ Symmetry of $AdS_3 \times S^3 \times S^3 \times S^1$ Superstrings,''
[arXiv:2306.13970 [hep-th]].

\bibitem{Ali:2024amc}
A.~Ali, M.~Ilahi, P.~P.~A.~Salih and S.~R.~Thottoli,
``Symmetry Structure of ADHM Sigma Models and $AdS_3$ Superstrings,''
[arXiv:2409.00474 [hep-th]].

\bibitem{Banados:1992wn}
M.~Banados, C.~Teitelboim and J.~Zanelli,
``The Black hole in three-dimensional space-time,''
Phys. Rev. Lett. \textbf{69}, 1849-1851 (1992)
[arXiv:hep-th/9204099 [hep-th]].

\bibitem{Ali:1992mj}
A.~Ali and A.~Kumar,
``O (d, d) transformations and 3-D black hole,''
Mod. Phys. Lett. A \textbf{8}, 2045-2052 (1993)
[arXiv:hep-th/9303032 [hep-th]].
 
\end{thebibliography}
\end{document}